\documentclass[journal]{IEEEtran}

\usepackage{graphicx,epsfig}
\usepackage{cite}
\usepackage{amsmath}
\usepackage{amssymb}
\usepackage{amsfonts,helvet}
\usepackage{footmisc}
\usepackage[alwaysadjust]{paralist}

\newtheorem{theorem}{Theorem}

\newtheorem{corollary}{Corollary}

\newtheorem{defn}{Definition}

\newtheorem{lemma}{Lemma}

\newtheorem{remark}{Remark}

\newtheorem{assumption}{\textbf{AS}}

\def\bE{{\mathbf{E}}}

\def\cN{\mbox{$\mathcal{N}$}}

\def\cR{\mbox{$\mathcal{R}$}}

\begin{document}
\title{Uplink Capacity and Interference Avoidance for Two-Tier Femtocell Networks}

\author{\authorblockN{Vikram Chandrasekhar, ~\IEEEmembership{Student~Member,~IEEE} and Jeffrey G. Andrews, ~\IEEEmembership{Senior~Member,~IEEE}} \\
\thanks{This research has been supported by Texas Instruments.
The authors are with the Wireless Networking
and Communications Group, Dept. of Electrical and Computer
Engineering at the University of Texas at Austin, TX 78712-1157.
(email:cvikram@mail.utexas.edu,jandrews@ece.utexas.edu)}}


\maketitle

\begin{abstract}
Two-tier femtocell networks-- comprising a conventional cellular
network plus embedded femtocell hotspots-- offer an economically
viable solution to achieving high cellular user capacity and
improved coverage. With universal frequency reuse and DS-CDMA
transmission however, the ensuing cross-tier interference
causes unacceptable outage probability. This paper develops an
uplink capacity analysis and interference avoidance strategy in such
a two-tier CDMA network. We evaluate a network-wide area spectral
efficiency metric called the \emph{operating contour (OC)} defined
as the feasible combinations of the average number of active
macrocell users and femtocell base stations (BS) per cell-site that
satisfy a target outage constraint. The capacity analysis provides
an accurate characterization of the uplink outage probability,
accounting for power control, path loss and shadowing effects.
Considering worst case interference at a corner femtocell, results reveal
that interference avoidance through a time-hopped CDMA physical
layer and sectorized antennas allows about a 7x higher femtocell
density, relative to a split spectrum two-tier network with
omnidirectional femtocell antennas. A femtocell exclusion region and
a tier selection based handoff policy offers modest improvements in
the OCs. These results provide guidelines for the design of robust
shared spectrum two-tier networks.
\end{abstract}


\section{Introduction}
\begin{figure} [!h]
\begin{center}
   \includegraphics[width=3.0in]{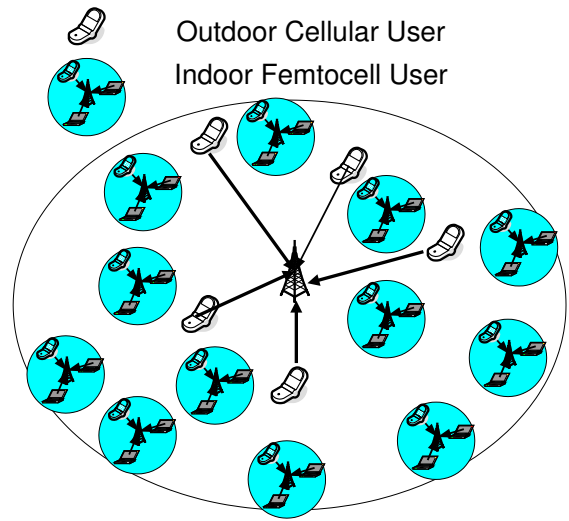}
   \caption{A Two-tier femtocell network}
   \label{fig:Twotiergnric}
   \end{center}
\end{figure}

Two-tier femtocell networks (Fig. \ref{fig:Twotiergnric})
are in the process of being deployed to improve cellular capacity
\cite{Shapira1994,Doufexi2003}. A femtocell serves as a small range
data access point situated around high user density hot-spots
serving stationary or low-mobility users. Examples of femtocells
include residential areas with home LAN access points, which are
deployed by end users and urban hot-spot data access points. A
femtocell is modeled as consisting of a randomly distributed
population of actively transmitting users. The femtocell radio range
($10-50$ meters) is much smaller than the macrocell radius
($300-2000$ meters) \cite{Ganz1997}. Users transmitting to
femtocells experience superior signal reception and lower their
transmit power, consequently prolonging battery life. The
implication is that femtocell users cause less interference to neighboring
femtocells and other macrocell users. Additionally, a two-tier
network offsets the burden on the macrocell BS, provided femtocells
are judiciously placed in traffic hot-spots, improving network
capacity and QoS.

Because of the scarce availability of spectrum and for reasons of flexible deployment, it may be easier for cellular operators to implement a two-tier network by sharing spectrum, rather than splitting spectrum between tiers.
The focus of this work is to answer the following questions:
\begin{itemize}
\item What is the two-tier uplink capacity in a typical
macrocell with randomly scattered hotspots, assuming a randomly
distributed population of actively transmitting users per femtocell?
\item Is it possible to accurately characterize the statistics of the cross-tier interference?
What is the effect of the femtocell hotspot density,
macrocell-femtocell power ratio and femtocell size?
\item How much benefit is accrued by interference avoidance using
antenna sectoring and time hopping in CDMA transmission? What is the
impact of using a femtocell exclusion region and a tier selection
policy for femtocell handoff?
\end{itemize}
By addressing these questions, our work augments existing research
on capacity analysis and interference mitigation in two-tier networks. We
show that creating a suitable infrastructure for curbing cross-tier
interference can actually increase the uplink capacity for a shared spectrum
network.

\subsection{Related work}
From a physical layer viewpoint, prior research has mainly focused
on analyzing the uplink capacity, assuming either a single
microcell\footnote{In the context of this paper, a microcell has a
much larger radio range (100-500 m) than a femtocell.} or multiple
regularly spaced microcells in a macrocell site. This model has
assumed significance for its analytical tractability, nonetheless,
it has limited applicability owing to the inherent variability in
microcell locations in realistic scenarios.

The ideas presented in this paper are most closely related to the
work by Kishore \emph{et al.} The downlink cellular capacity of a
two-tier network is derived in \cite{Kishore2003DL}. The results
show that the cellular user capacity is limited by uplink
performance for both slow and fast power control. In
\cite{Kishore2003}, the OCs for a two-tier network are derived for
different tier-selection schemes, assuming an arbitrarily placed
microcell. Further work by the same author
\cite{Kishore2006a,Kishore2006} extended the framework to multiple
microcells embedded inside multiple macrocells. The cross-tier interference
is approximated by its average and cross-tier microcell to microcell
interference is ignored. The resulting analysis is shown to be accurate only
up to 8 microcells per macrocell. Our results, on the other hand, are
accurate over a wide range of femtocell densities, without
approximating the interference statistics.

Related work includes \cite{Kim2000}, which discusses the benefits
of having a tilted antenna radiation pattern and macrocell-microcell
power ratio control. In \cite{Wu1997,
Karlsson1999}, a regular network comprising a large hexagonal
macrocell and smaller hexagonal microcells is considered. Citing
near far effects, the authors conclude that it is more practical to
split the RF spectrum between each tier. The reason being that the
loss in trunking efficiency by splitting the spectrum is lower than
the increase in outage probability in a shared spectrum two-tier
network. Our paper, in contrast, shows a higher user capacity for a
shared spectrum network by enforcing \emph{higher spatial reuse}
through small femtocells and \emph{interference avoidance} by way of
antenna sectoring and Time hopped CDMA (TH-CDMA) in each tier.

Finally, from a network perspective, Joseph \emph{et al.}
\cite{Joseph2004} study impact of user behavior, load
balancing and different pricing schemes for interoperability between
Wi-Fi hotspots and cellular networks. In \cite{Ganz1997}, the design
of a multitiered wireless network with Poisson call arrivals is
formulated as an constrained optimization problem, and the results
highlight the economic benefits of a two-tier network
infrastructure: increased stability in system cost and
a more gradual performance degradation as users are added.

\subsection{Contributions}
This paper employs a stochastic geometry framework for modeling the
\emph{random} spatial distribution of users/femtocells, in contrast
to prior work \cite{Kishore2003, Kishore2006,
Kishore2006a, I1993, Wu1997, Karlsson1999}. Hotspot locations are
likely to vary from one cellsite to another, and be opportunistic
rather than planned: Therefore a capacity analysis that embraces
instead of neglecting randomness will naturally provide more
accurate results and more plausible insights.

To model the user/hotspot locations, the paper assumes that the
macrocell users and femtocell BS are randomly distributed as a
Homogeneous Spatial Poisson Point Process (SPPP) (see \cite{Kingman,Haenggi2009} for background, prior works include \cite{Chan2001,Baccelli2001,Foss1996}).
The three key contributions in our paper are summarized below.
\begin{asparaitem}
\item First, a novel outage probability analysis is presented, accounting
for cellular geometry, cross-tier interference and
shadowing effects. We derive tight lower bounds on statistics
of macrocell interference at any femtocell hotspot BS along the hexagonal
axis. Next, assuming small femtocell sizes, a Poisson-Gaussian model
for macrocell interference and alpha-stable distribution for cross-tier
femtocell interference is shown to accurately capture the statistics at the
macrocell BS. In the analysis, outage events are explicitly modeled
rather than considering average interference as in
\cite{Wu1997,I1993}. For doing so, the properties of Poisson
shot-noise processes (SNP)\cite{Lowen1990,Ilow1998} and Poisson void
probabilities \cite{Kingman} are used for deriving the uplink outage
probabilities.

\item Second, robust interference avoidance is shown to enable
two-tier networks with universal frequency reuse to achieve higher
user capacity, relative to splitting the spectrum across tiers. With interference
avoidance, an equitable distribution of users between tier 1 and
tier 2 networks is shown to be achieved with an \emph{order-wise
difference} in the ratio of their received powers. Even considering
the worst case cross-tier interference at a corner femtocell, results for
moderately loaded cellular networks reveal that interference
avoidance provides a $7$x increase in the mean number of femtocells over
split spectrum two-tier networks.

\item Third, additional interference avoidance using a combination of femtocell exclusion and
tier selection based femtocell handoff offers modest improvements in the network OCs. This suggests
that at least for small femtocell sizes, time hopping and antenna sectoring offer the
largest gains in user capacity for shared spectrum two-tier networks.
\end{asparaitem}
%

\section{System Model}
\label{Se:Sysmod} Denote $\mathcal{H} \subset \mathbb{R}^2$ as the interior
of a reference hexagonal macrocell $C$
of radius $R_c$. The tier 1 network consists of low density
cellular users that are communicating with the central BS in
each cellsite. The cellular users are distributed on
${\mathbb{R}}^{2}$ according to a homogeneous SPPP $\Omega_c$ of
intensity $\lambda_c$. The overlaid tier 2 network containing the
femtocell BS's forms a homogeneous SPPP\footnote{The system model
allows a cellular user to be present inside a femtocell as the
governing process $\Omega_c$ is homogeneous.} $\Omega_{f}$ with
intensity $\lambda_f$. Each femtocell hotspot includes a Poisson
distributed population of actively transmitting users\footnote{A
hard handoff is assumed to allocate subscribed hotspot users to a femtocell, provided
they fall within its radio range.} with mean $U_f$ in a circular coverage area of radius
$R_f, R_f \ll R_c$. To maximize user capacity per cellsite, it is
desirable to have $\lambda_f \gg \lambda_c$; as will be shown,
cross-tier interference at a macrocell BS limits $\lambda_f$ for a given
$\lambda_c$. Defining $|\mathcal{H}| \triangleq 2.6R_c^{2}$ as the area of
the hexagonal region $\mathcal{H}$, the mean number of macrocell users and
femtocell BS's per cellsite are given as $N_c=\lambda_c \cdot |\mathcal{H}|$
and $N_f=\lambda_f \cdot |\mathcal{H}|$ respectively. Table
\ref{Tbl:Sysprms} shows a summary of important parameters and
typical values for them, which are used later in numerical
simulations.

Users in each tier employ DS-CDMA with processing gain $G$.
Uplink power control adjusts for propagation losses and
log-normal shadowing, which is standard in contemporary CDMA
networks. The macrocell and femtocell receive powers are denoted as
$P_r^{c}$ and $P_r^{f}$ respectively. Any power control errors \cite{Jansen1995}
and short-term fading effects are ignored for analytical convenience.
We affirm this assumption as reasonable, especially in a wideband system with significant
frequency diversity and robust reception (through RAKE receiver,
coding and interleaving).

\subsection{TH-CDMA and Antenna sectoring}
 Suppose that the CDMA period $T=G \cdot T_c$ is divided into $N_{\textrm{hop}}$
hopping slots, each of duration $T/N_{\textrm{hop}}$. Every macrocell user
and femtocell (all active users within a femtocell transmit in the
same hopping slot) independently choose to transmit over any one
slot, and remain silent over the remaining $N_{\textrm{hop}}-1$ slots. The
resulting intra- and cross-tier interference are ``thinned'' by a
factor of $N_{\textrm{hop}}$ \cite{Kingman}. Using TH-CDMA, users in each
tier effectively sacrifice a factor $N_{\textrm{hop}}$ of their processing
gain, but benefit by thinning the interfering field by the same
factor.

We further assume sectorized antenna reception in both the macrocell and femtocell BS, with
antenna alignment angle $\theta$ and sector width equaling
$2\pi/N_{\textrm{sec}}$. While antenna sectoring is a common feature at the
macrocell BS in practical cellular systems, this paper proposes to
use sectorized antennas at femtocell BS's as well. The reason is that
the cross-tier interference caused by nearby cellular users can lead to
unacceptable outage performance over the femtocell uplink; this
motivates the need for directed femtocell antennas. The spatial
thinning effect of TH-CDMA transmission and antenna sectoring is
analytically derived in the following lemma.
\begin{lemma} [Spatial thinning by interference avoidance]
\label{Le:Thinning} \emph{With TH-CDMA transmission over $N_{\textrm{hop}}$
slots and antenna sectoring with $N_{\textrm{sec}}$ directed BS antennas in
each tier, the interfering field at a given antenna sector can be
mapped to the SPPPs $\Phi_c$ and $\Phi_f$ on $\mathbb{R}^2$ with
intensities $\eta_c=\lambda_c/(N_{\textrm{hop}}N_{\textrm{sec}})$ and
$\eta_f=\lambda_f(1-e^{-U_f})/(N_{\textrm{hop}}N_{\textrm{sec}})$ respectively.
}
\end{lemma}
\begin{proof}
See Appendix \ref{Proof:Thinning}.
\end{proof}
The following definitions will be useful in the remainder of
the paper.
\begin{defn}
\label{De:def1} Denote $\mathcal{H}_{sec} \subseteq \mathcal{H}$ as the region
within $\mathcal{H}$ covered by a antenna sector corresponding to a
macrocell BS or a femtocell BS within the reference cellsite. For
example, $\mathcal{H}_{sec}=\mathcal{H}$ for an omnidirectional femtocell located at
the corner of the reference macrocell.
\end{defn}
\begin{defn}
\label{De:def2} Denote $\hat{\Omega}_c$ and $\hat{\Omega}_f$ as the
heterogeneous SPPPs composed of active macrocell and femtocell
interferers as seen at a antenna sector in each tier, whose
intensities are given by $\hat{\lambda}_c$ and $\hat{\lambda}_f$ in
\eqref{eq:sctrSPPP}. Denote the equivalent mapped homogeneous SPPPs
over $\mathbb{R}^{2}$ by $\Phi_c$ and $\Phi_f$ whose intensities are
given by $\eta_c$ and $\eta_f$ respectively.
\end{defn}
\begin{defn}
\label{De:def3} Denote the restriction of $\hat{\Omega}_c$ and
$\hat{\Omega}_f$ to $\mathcal{H}$ by the SPPPs $\Pi_c$ and $\Pi_f$
respectively.
\end{defn}

\subsection{Channel Model and Interference}
The channel is represented as a combination of path loss and
log-normal shadowing. The path loss exponents are denoted by
$\alpha$ (outdoor transmission) and $\beta$ (indoor femtocell
transmission) with random lognormal shadowing standard deviation $\sigma_{dB}$.
Through uplink power control, a macrocell user transmitting at a random
position $X$ w.r.t the reference macrocell BS $C$ chooses a transmit
power level $P_t^{c}=P_r^{c}/g_c(|X|)$. Here $g_c(|X|)$ is the
attenuation function for outdoor propagation, defined as
$g_c(|X|)=K_{c}(d_{0c}/|X|)^{\alpha}\Theta_{C}$ where
$10\log_{10}{\Theta_C} \sim \cN(0,\sigma_{dB}^2) $ is the log-normal
shadowing from user to $C$, $K_c \triangleq [c/(4\pi f_{c}
d_{0c})]^{2}$ is a unitless constant that depends on the wavelength
of the RF carrier $c/f_c$ and outdoor reference distance $d_{0c}$.
\begin{figure} [!h]
\begin{center}
       \includegraphics[width=2.0in]{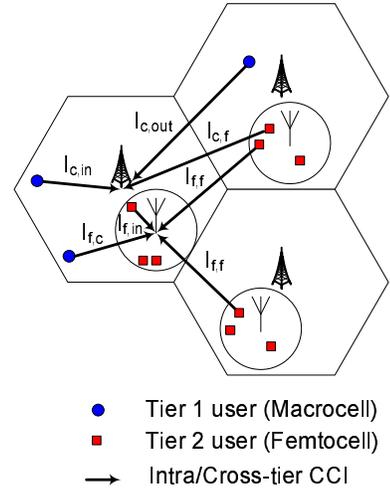}
       \caption{Intra-tier and cross-tier interference at each tier. The arrows denote the interference arising from either a Tier 1 or Tier 2 user.}
       \label{fig:Intrfmdlng}
   \end{center}
\end{figure}
Similarly, a femtocell user at a random position $Y$ within a femtocell BS $F$
chooses a transmit power $P_t^{f}=P_r^{f}/g_f(|Y|)$, where
$g_f(|Y|)=K_{f}(d_{0f}/|Y|)^{\beta}\Theta_{F}$ represents the indoor attenuation,
$10\log_{10}{\Theta_F} \sim \cN(0,\sigma_{dB}^2) $ and $K_f
\triangleq [c/(4\pi f_{c} d_{0f})]^{2}$. Here $d_{0f}$ is the
reference distance for calculating the indoor propagation loss. Note
that in reality, $K_{c}$ and $K_{f}$ are empirically determined. The
interference in each tier (Fig. \ref{fig:Intrfmdlng}) can be grouped
as:
\begin{asparadesc}
\item[Macrocell interference at a macrocell.]
Through power control, all macrocell users within $\mathcal{H}_{sec}$ are
received with constant power $P_r^{c}$, so the in-cell interference
equals $(N-1)P_r^{c}$, where $N \sim
\mathrm{Poisson}(N_c/N_{\textrm{hop}})$. As such, inferring the exact
statistics of out-of-cell cellular interference $I_{c,\textrm{out}}$ is
analytically intractable; it is assumed that $I_{c,\textrm{out}}$ is
distributed according to a scaled Gaussian pdf \cite{Chan2001}.
Defining $\mu$ and $\sigma^2$ to be the empirically determined
parameters of the Gaussian, the pdf of $I_{c,\textrm{out}}$ is given as $
f_{I_{c,\textrm{out}}}(y)=\frac{2e^{-\frac{1}{2}(y-\mu)^2/ \sigma^2}}{\sqrt{2\pi\sigma^2}\lbrack 2-
\textrm{erfc}(\frac{\mu} {\sqrt{2}\sigma}) \rbrack}$, where $\textrm{erfc}(t)
\triangleq \sqrt{\frac{2}{\pi}}\int_{t \sqrt{2}}^{\infty} e^{-x^2/2} dx$.
\item[Femtocell interference at a macrocell.]
Say femtocell $F_{i}$ with $U_i \sim \mathrm{Poisson}(U_f)$ users is
located at random position $X_i$ w.r.t the reference macrocell BS $C$.
Inside $F_i$, a randomly placed Tier 2 user $j$ at distance $Y_j$
from the femtocell BS transmits with power
$P_t^{f}(j)=P_r^{f}/g_f(Y_j)$. The interference caused at $C$ from
user $j$ inside $F_i$ is given as,
\begin{align}
\label{eq:maccrossfemto}
I_{c,f}(F_i,j)&=P_r^{f}g_c(|X_i+Y_j|)/g_f(|Y_j|) \notag \\
&\approx P_r^{f}g_c(|X_i|)/g_f(R_f) \notag \\
&=Q_f \Theta_{j,C}/\Theta_{j,F_i}|X_i|^{-\alpha}
\end{align} where
$Q_f \triangleq
P_r^{f}R_f^{\beta}(\frac{K_{c}}{K_{f}})(\frac{d_{0c}^{\alpha}}{d_{0f}^{\beta}})$.
In doing so, we make two important assumptions:
\begin{assumption}
For small sized femtocells ($R_f \ll R_c$), a
femtocell or macrocell BS sees interference from other femtocells
as a \emph{point source} of interference, implying $g_c(|X_i+Y_j|)
\approx g_c(|X_i|)$.
\end{assumption}
\begin{assumption}
When analyzing the interference caused by a random femtocell $F_i$
at any other location, the $U_i$ femtocell users can be modeled as
transmitting with maximum power, so that $g_f(|Y_j|) \approx
g_f(R_f)$. This is for analytical tractability and modeling
worst-case interference.
\end{assumption}

Summing \eqref{eq:maccrossfemto} over all femtocells over a antenna
sector at a macrocell BS, the cumulative cross-tier interference at the
reference macrocell BS $C$ is represented by the Poisson SNP
\cite{Lowen1990},
\begin{align}
\label{eq:PoisSNP}
 I_{c,f}=\sum_{F_{i} \in \hat{\Omega}_{f}} Q_f
\Psi_i|X_i|^{-\alpha}
\end{align} where $\Psi_i \triangleq
\sum_{l=1}^{U_i}\Theta_{l,C}/\Theta_{l,F_i}$ defines the cumulative
shadowing gain between actively transmitting users in femtocell
$F_i$ and macrocell BS $C$.
\item [Neighboring femtocell interference at a femtocell.]
By an identical argument as above, the interference caused at the BS
antenna sector of femtocell $F_{j}$ from other femtocells $F_i, i
\neq j$ is a Poisson SNP given by $I_{f,f}=\sum_{F_{i} \in
\hat{\Omega}_{f}} Q_f \Psi_i |X_i|^{-\alpha}$, where $|X_i|$ refers
to the distance between $(F_i,F_j)$ and $\Psi_i \triangleq
\sum_{l=1}^{U}\Theta_{l,F_j} / \Theta_{l,F_i}$.
\item [Interference from active users within a femtocell.]
Conditioned on the femtocell containing $U$ actively
transmitting users ($U \geq 1$), the intra-tier interference experienced by the user of
interest arising from simultaneous transmissions within the
femtocell is given as $I_{f,\textrm{in}}=(U-1)
P_r^{f},\mathbb{E}[U]=\frac{U_f}{1-e^{-U_f}}$.
\item [Macrocell interference at a femtocell.]
This paper analyzes outage probability at a femtocell BS $F_j$
located on the hexagonal axis, considering the effect of in-cell
cellular interference. The cumulative tier 1 interference $I_{f,c}$ experienced at a femtocell
can be lower bounded by the interference caused by the set of tier 1 interferers inside the
reference macrocell $\Pi_c$. This lower bound is represented as $I_{f,c} \geq I_{f,c}^{lb}=\sum_{i \in
\Pi_c}P_r^{c}\Psi_i(\frac{|X_i|}{|Y_i|})^{\alpha}$, where $\Psi_i
\triangleq \Theta_{i,F_j}/\Theta_{i,C},10\log_{10}\Psi_i \sim
\cN(0,2\sigma_{dB}^2)$ is the LN shadowing term and $|X_i|, |Y_i|$
represent the distances of macrocell user $i$ to the macrocell BS
and femtocell BS respectively. Observe that a corner femtocell
experiences a significantly higher macrocell interference relative to an
interior femtocell, therefore the cdf $F_{I_{f,c}}(\cdot)$ is not a
stationary distribution.
\end{asparadesc}

\section{Per Tier Outage Probability}
\label{Se:OutageProb} To derive the OCs, an uplink outage
probability constraint is formulated in each tier. Define $N_f$ and
$N_c$ as the average number of femtocell BS's and macrocell users
per cellsite respectively. A user experiences outage if the instantaneous received
Signal-to-Interference Ratio (SIR) over a transmission
is below a threshold $\gamma$. Any feasible
$(\tilde{N_f},\tilde{N_c})$ satisfies the outage probability
requirements $\mathbb{P}_{\textrm{out}}^{f} \leq
\epsilon,\mathbb{P}_{\textrm{out}}^{c} \leq \epsilon$ in each tier. The
outage probabilities $\mathbb{P}_{\textrm{out}}^{c}(N_f,N_c)$ [resp.
$\mathbb{P}_{\textrm{out}}^{f}(N_f,N_c)$] are defined as the probabilities
that the despread narrowband SIR for a macrocell user
[femtocell user] at the tier 1 [tier 2] antenna sector is below $\gamma$.
Assuming the PN code cross-correlation equals
$N_{\textrm{hop}}/G$\footnote{With $N_{\textrm{hop}}=G=1$, the model reduces to a non
CDMA narrowband transmission; with $N_{\textrm{hop}}=G \gg 1$, the model
reduces to a timeslotted ALOHA transmission}, define
\begin{align}
\label{eq:Outg_const} \mathbb{P}_{\textrm{out}}^{c}(N_f,N_c)&=\mathbb{P}
\Biggl(\frac{G/N_{\textrm{hop}}P_r^{c}}{I_{c,\textrm{in}}+I_{c,\textrm{out}}+I_{c,f}} \leq
\gamma \Big \vert \lvert \hat{\Omega}_c \rvert \geq 1 \Biggr) \notag \\
\mathbb{P}_{\textrm{out}}^{f}(N_f,N_c)&=\mathbb{P}\Biggl(\frac{G/N_{\textrm{hop}}P_r^{f}}{(U-1)
\cdot P_r^{f}+I_{f,f}+I_{f,c}} \leq \gamma \Big \vert U \geq 1
\Biggr)
\end{align}
where $|\hat{\Omega}_c|$ denotes the number of points in
$\hat{\Omega}_c$ and the unconditioned $U \sim
\mathrm{Poisson}(U_f/N_{\textrm{sec}})$.
 The OCs for the macrocell [resp.
femtocell] are obtained by computing the Pareto optimal $(N_f,N_c)$ pairs which satisfy a target outage
constraint $\epsilon$. More formally,
\begin{align}
\label{Ch2:eq:Twotier_OC}
(\tilde{N_f},\tilde{N_c}) = \lbrace (N_f,N_c)&: \not\exists (N_f' > N_f ,N_c' > N_c), \\
     &\mathbb{P}_{\textrm{out}}^{c}(N_f',N_c') \leq \epsilon, \mathbb{P}_{\textrm{out}}^{f}(N_f',N_c') \leq \epsilon \rbrace
\end{align}
The OCs for the two-tier network are obtained corresponding to those
feasible combinations of $(\tilde{N}_c,\tilde{N}_f)$ that
simultaneously satisfy $\mathbb{P}_{\textrm{out}}^{f}\leq\epsilon$ and
$\mathbb{P}_{\textrm{out}}^{c}\leq\epsilon$ respectively. For doing so, we
derive the following theorems which quantify the outage
probabilities and interference statistics in each tier.
\begin{theorem}
\label{Th:Crossint_fem} \emph{ For small femtocell sizes, the
statistics of the cross-tier femtocell interference $I_{c,f}$ (and intra-tier
femtocell interference $I_{f,f}$) at a antenna sector are given by a
Poisson SNP $Y=\sum_{i \in \Phi_{f}}Q_f \Psi_i |X_i|^{-\alpha}$ with
iid $\Psi_{i}=\sum_{j=1}^{U_i}\Psi_{ij}, 10\log_{10} \Psi_{ij} \sim
\cN(0,\sigma_{dB}^2), U_i \sim U|U \geq 1$ and $U \sim
\mathrm{Poisson}(U_f)$. In particular, if the outdoor path loss
exponent $\alpha=4$, then $Y$ follows a L\'{e}vy-stable distribution
with stability exponent $1/2$, whose probability density function (pdf) and cumulative distribution function (cdf) are given as,
\begin{align}
\label{eq:DF1DSPPP}
f_Y(y)=\sqrt{\frac{\kappa_f}{\pi}}y^{-3/2}e^{-\kappa_f/y},
{F}_Y(y)=\mathrm{erfc}\Biggl(\sqrt{\frac{\kappa_f}{y}}\Biggr)
\end{align}
where $\kappa_f \triangleq \eta_f^2 {\pi}^3 Q_f
(\bE[\Psi^{1/2}])^2/4$.}
\end{theorem}
\begin{proof}
See Appendix \ref{Proof:Crossint_fem}.
\end{proof}

\begin{figure} [!h]
\begin{center}
   \includegraphics[width=3.5in]{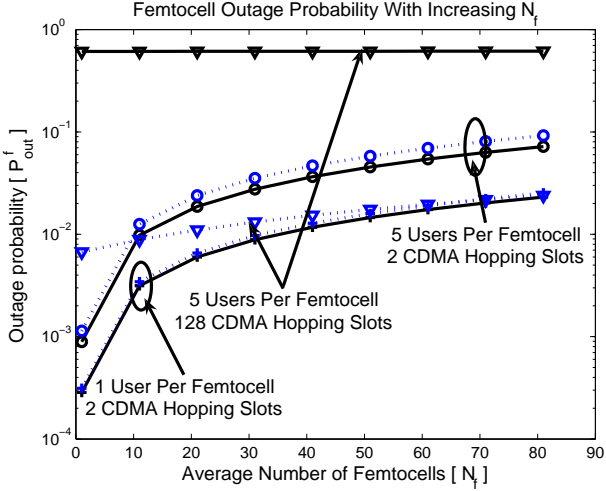}
   \caption{Comparison of joint and independent hopping protocols at a femtocell BS with sectorized antennas. Solid lines represent the joint hopping performance when all users within a femtocell share a common hopping slot. Dotted lines represent the performance when every femtocell user is assigned an independent CDMA hopping slot. }
   \label{fig:ShPl_Mic_HoppingProt_Nhop}
   \end{center}
\end{figure}

\begin{remark} [Femtocell size]
Increasing femtocell size ($R_f$) \emph{strictly increases} the
outage probabilities arising from the femtocell interference $I_{f,f}$ and
$I_{c,f}$ in a two-tier network. To elucidate this, observe that an
increase in $R_f$ causes $\kappa_f$ to increase by a factor
$R_f^{\beta}$. By monotonicity of $\mathrm{erfc}(\cdot)$, the cdf's
$F_{I_{f,f}}(\cdot),F_{I_{c,f}}(\cdot)$ decrease as $\kappa_f$
increases, causing a higher outage probability per tier.
Intuitively, a femtocell user located on the edge of a femtocell
will cause excessive interference at a nearby femtocell BS; this edge effect
appears as a power control factor $R_f^{\beta}$ in
\eqref{eq:DF1DSPPP}.
\end{remark}
\begin{remark} [Hopping Protocol]
All Tier 2 users within a femtocell are assumed to jointly
choose a hopping slot. Suppose we compare this against an independent hopping protocol,
where users within a femtocell are independently assigned a hopping slot.
With independent hopping, the intensity of $\Phi_f$ equals
$\tilde{\eta}_f=\frac{\lambda_f}{N_{\textrm{sec}}} \cdot
(1-e^{-U_f/N_{\textrm{hop}}})$ (note the difference from $\eta_f$ in Lemma
\ref{Le:Thinning}) and the average number of interfering users in an
actively transmitting femtocell equals
$\frac{U_f/N_{\textrm{hop}}}{1-e^{-U_f/N_{\textrm{hop}}}}$. With an outage threshold
of $P_r^{f}G/(\gamma N_{\textrm{hop}})$ \eqref{eq:Outg_const} at a femtocell
BS, two observations are in order:
\begin{asparadesc}
\item[TH-CDMA transmission:] When $\frac{G}{N_{\textrm{hop}}} \gg 1$, joint hopping
is preferable from an outage probability perspective. Intuitively,
joint hopping reduces $\lambda_f$ by a factor $N_{\textrm{hop}}$, causing a
quadratic decrease in $\kappa_f$ in \eqref{eq:DF1DSPPP}; independent
hopping decreases the number of interfering users per active
femtocell, causing a sub-quadratic decrease in $\mathbb{E}\lbrack
{\Psi}^{1/2} \rbrack^2$. The consequence is that joint hopping
results in a greater decrease in $\mathbb{P}_{\textrm{out}}^{f}$. Using
$N_{\textrm{hop}}=2$, Fig. \ref{fig:ShPl_Mic_HoppingProt_Nhop} confirms this
intuition; notably, the gap in outage performance is dictated by the
hotspot user density: In heavily loaded femtocells ($U_f \gg
1$), a joint hopping scheme is clearly superior. For lightly
loaded femtocells, $\eta_f \simeq \tilde{\eta}_f \approx
\frac{\lambda_f U_f}{N_{\textrm{sec}} \cdot N_{\textrm{hop}}}$, implying that
independent and joint hopping schemes perform nearly identical.
\item[Random Access transmission:] When $N_{\textrm{hop}}=G \gg 1$, the femtocell
outage threshold is $P_r^{f}/\gamma$; by consequence, it is
preferable to use independent hopping across the tier 2 network (see
Fig. \ref{fig:ShPl_Mic_HoppingProt_Nhop}). With joint hopping, even a
\emph{single interferer} within a femtocell can cause outage for the user
of interest as there is no CDMA interference averaging; in contrast, independent hopping offers
increased interference avoidance since the likelihood of two
femtocell users sharing a hopping slot is negligible. Consequently,
in non-CDMA two-tier cellular networks employing interference
avoidance, an independent assignment of hopping slots is
preferable from an outage viewpoint.
\end{asparadesc}
\end{remark}
Using Theorem \ref{Th:Crossint_fem}, the cellular outage
probability is now formulated.
\begin{theorem}[Macrocell outage probability]
\label{Th:Pout_mac}\emph{Let the outdoor path loss exponent $\alpha=4$.
With Poisson in-cell macrocell interference $I_{c,\textrm{in}}$, Gaussian out-of-cell
interference $I_{c,\textrm{out}}$ and L\'{e}vy-stable femtocell interference $I_{c,f}$ given by
\eqref{eq:DF1DSPPP}, the outage probability at the macrocell BS
antenna sector is given as,
\begin{align}
\epsilon \geq \mathbb{P}_{\textrm{out}}^{c}=1-\frac{1}{1-e^{-\eta_c
|\mathcal{H}|}}\sum_{m=1}^{\lfloor \rho_c/P_r^{c} \rfloor}\frac{e^{-\eta_c
|\mathcal{H}|}{(\eta_c |\mathcal{H}|)^m}}{m!}G_c(\tilde{\rho}_c)
\end{align}
where $\eta_c=\frac{\lambda_c}{N_{\textrm{hop}}N_{\textrm{sec}}}$,$\rho_c=\frac{P_r^{c}G}{\gamma N_{\textrm{hop}}}, \tilde{\rho}_c=\rho_c-(m-1)P_r^{c}$ and
$G_c(t) \triangleq \int_{0}^{t}f_{I_{c,\textrm{out}}}(t-y)F_{I_{c,f}}(y) dy$
}
\end{theorem}
\begin{proof}
See Appendix \ref{Proof:Pout_mac}.
\end{proof}
Theorems \ref{Th:Crossint_fem} and \ref{Th:Pout_mac} provide the
tools to quantify the largest $N_f$ that can be accommodated at a
given $N_c$ subject to an outage constraint $\epsilon$. The next
step is to compute the outage probability at a femtocell as defined
in \eqref{eq:Outg_const}. To do so, assume that the femtocell is
located on the axis at a distance $R_0$ from the macrocell center
and the femtocell antenna sector is aligned at angle
$\theta$ w.r.t the hexagonal axis.

\begin{figure} [!h]
\begin{center}
       \includegraphics[width=3.5in]{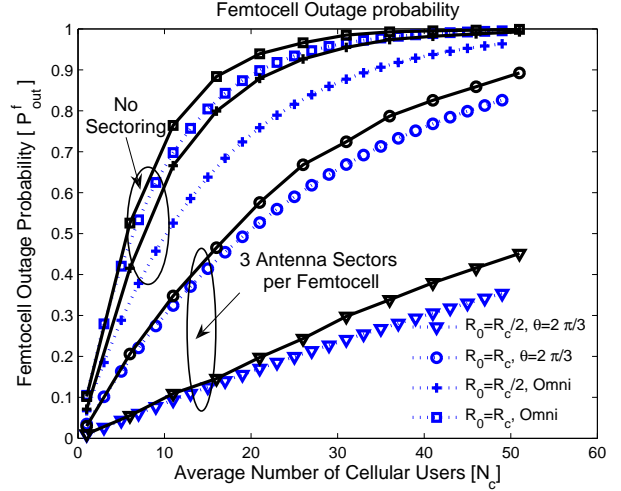}
       \caption{Lower bounds on outage probability for an interior and corner Femtocell ($N_{\textrm{hop}}=1, P_r^{f}=P_r^{c}$). Dotted lines indicate the theoretical lower bounds on outage probability while solid lines indicate the empirically estimated probabilities.}
       \label{fig:Pout_Fem_CrossMacroIntrf}
   \end{center}
\end{figure}

The following theorem derives a lower bound on the tail probability
for the distribution of the tier 1 interference $I_{f,c}$,
experienced at any femtocell located along the hexagonal axis.
\begin{theorem} [Lower bound on cellular interference]
\label{Th:Crossint_mac} \emph{At any femtocell antenna sector
located at distance $0 < R_0 \leq R_c$ from the macrocell BS along
the hexagonal axis:
\begin{enumerate}
\item The complementary cumulative distribution function (ccdf) of the cellular interference $I_{f,c}$ over a femtocell
antenna sector is lower bounded as $\bar{F}_{I_{f,c}}(y)\geq
1-F^{lb}_{I_{f,c}}(y)$, where:
\begin{align} \label{eq:Pout_sctrdintfemlb}
 F^{lb}_{I_{f,c}}(y) & = \exp \left \{-\frac{\lambda_c}{N_{\textrm{hop}}} \iint \limits_{\mathcal{H}_{sec}}
S(r,\phi;y) r dr d\phi \right\} \\
S(r,\phi;y) &\triangleq \bar{F}_{\Psi}\lbrack y/P_r^{c}\cdot
(r/|re^{\mathbf{i}\phi}+R_0|)^{\alpha}\rbrack \notag
\end{align}
Here $\bar{F}_{\Psi}$ is the ccdf
of $\Psi: 10\log_{10}\Psi \sim \cN(0,2\sigma_{dB}^2)$, $\mathbf{i} \triangleq \sqrt{-1}$, $\theta$ is
the femtocell BS antenna alignment angle and $\mathcal{H}_{sec} \subseteq
\mathcal{H}$ denotes the region inside the reference macrocell enclosed
between $\theta \leq \phi \leq \theta+2\pi/N_{\textrm{sec}}$.
 \item For a corner femtocell $R_0=R_c$ with an omnidirectional femtocell antenna
 $N_{\textrm{sec}}=1$, the ccdf of $I_{f,c}$ is lower bounded as
$\bar{F}_{I_{f,c}}(y)\geq 1-F^{lb}_{I_{f,c}}(y)$, where :
\begin{align}
\label{eq:Pout_omnicorfemlb} F^{lb}_{I_{f,c}}(y) = \exp \left
\{-3\frac{\lambda_c}{N_{\textrm{hop}}}\iint\limits_{\mathcal{H}}S(r,\phi;y)r \textrm{d}r \textrm{d}\phi
\right\}
\end{align}
\end{enumerate}
 }
\end{theorem}
\begin{proof}
See Appendix \ref{Proof:Crossint_mac}.
\end{proof}
For a path loss only model, the lower bounds on the femtocell outage
probability  can be derived analogously as stated in the following corollary.
\begin{corollary}
\emph{With the above definitions, assuming a pure path loss model
(no shadowing), \eqref{eq:Pout_sctrdintfemlb} and
\eqref{eq:Pout_omnicorfemlb} hold with $S(r,\phi;y) \triangleq
\mathbf{1}\lbrack P_r^{c} \cdot (|re^{\mathbf{i}\phi}+R_0| / r)^{\alpha} \geq
y \rbrack $}
\end{corollary}
Theorem \ref{Th:Crossint_mac} characterizes the relationship between
the intensity of macrocell users and the femtocell outage
probability. Observe that the outage probability
$\bar{F}^{lb}_{I_{f,c}} \rightarrow 1$ exponentially, as
$\lambda_c \rightarrow \infty$. Further, increasing $N_{\textrm{hop}}$
``thins'' the intensity of $\Pi_c$, thereby mitigating cross-tier
interference at the femtocell BS. Fig. \ref{fig:Pout_Fem_CrossMacroIntrf}
depicts the outage lower bounds to evaluate the impact of
cellular interference $I_{f,c}$. Corresponding to an interior and corner
femtocell location, the lower bounds are computed when the femtocell
BS antenna is either sectorized-- $N_{\textrm{sec}}=3$ with antenna alignment
angle $\theta=2\pi/3$ -- or omnidirectional. No hopping is used
($N_{\textrm{hop}}=1$), while a unity power ratio ($P_r^{f}/P_r^{c}=1$) is
maintained. Two observations are in order:
\begin{asparadesc}
\item [Tightness of lower bound:]
The tightness of \eqref{eq:Pout_sctrdintfemlb} and
\eqref{eq:Pout_omnicorfemlb} shows that the cross-tier interference $I_{f,c}$
is primarily impacted by the set of \emph{dominant cellular
interferers} \eqref{eq:SPPPartng}. The implication is that one can
perform accurate outage analysis at a femtocell by considering only
the cellular users whose transmissions are strong enough to individually cause outage. This
agrees with the observations in \cite{Weber2005,Weber2007}.
\item [Infeasibility of omnidirectional femtocells:]
The benefits of sectorized antennas for interference mitigation at the
femtocell BS are evident; with a sectorized BS antenna, a corner
femtocell (worst-case macrocell interference) performs considerably better
than an interior omnidirectional femtocell.
\end{asparadesc}
Using Theorems \ref{Th:Crossint_fem} and \ref{Th:Crossint_mac}, the
femtocell outage probability \eqref{eq:Outg_const} is stated in the
next theorem.
\begin{theorem} [Femtocell outage probability]
\label{Th:Pout_fem}\emph{ Let outdoor path loss exponent $\alpha=4$.
For small $\lambda_c$, the femtocell outage probability
$\mathbb{P}_{\textrm{out}}^{f}$ is lower bounded as:
\begin{align}
\epsilon \geq \mathbb{P}_{out}^{f,lb} \approx
1-\frac{e^{-U_{f,\textrm{sec}}}}{1-e^{-U_{f,\textrm{sec}}}}\sum_{m=1}^{\lfloor
\rho_f/P_r^{f}\rfloor}\frac{U_{f,\textrm{sec}}^m}{m!} \cdot
G_{f}(\tilde{\rho}_f)
\end{align}
where
$U_{f,\textrm{sec}} \triangleq \frac{U_f}{N_{\textrm{sec}}}, \rho_f \triangleq
\frac{G P_r^{f}}{N_{\textrm{hop}} \cdot \gamma}$,
$\tilde{\rho}_f=\rho_f-(m-1)\cdot P_r^{f}$ and $G_f(t) \triangleq
F_{I_{f,f}}(t)+\int_{0}^{t}f_{I_{f,f}}(t-y)\ln{(F^{lb}_{I_{f,c}}(y))}
dy$. }
\end{theorem}
\begin{proof}
See Appendix \ref{Proof:Pout_fem}.
\end{proof}
For a given $N_f$, Theorem \ref{Th:Pout_fem} computes the
largest $N_c$ which ensures the SIR threshold $\gamma$ is
satisfies for a fraction ($1-\epsilon$) of the time. Furthermore,
the lower bound $F^{lb}_{I_{f,c}}(\cdot)$ was shown to be tight,
hence the computed $N_c$ is not overly optimistic. Using Theorems
\ref{Th:Pout_mac} and \ref{Th:Pout_fem}, the OCs for the two-tier
network with interference avoidance can now be readily obtained. The
following section studies using a femtocell exclusion region around
the macrocell BS and a tier selection based femtocell handoff
policy, in addition to the interference avoidance strategies
discussed hitherto.

\section{Femtocell exclusion region and Tier Selection}
Suppose the reference macrocell BS has a femtocell exclusion region
${\cR}_{f,\textrm{exc}} \subset \mathcal{H}$ surrounding it. This idea is motivated
by the need to silence neighboring femtocell transmissions which are
strong enough to individually cause outage at a macrocell BS;
similar schemes have been proposed in \cite{Hasan2007} and adopted
in the CSMA scheduler in the 802.11 standard. The tier 2 femtocell
network then forms a heterogeneous SPPP on $\mathcal{H}$ with the average
number of femtocells in each cell-site equaling $\lambda_f \cdot
(|\mathcal{H}|-|{\cR}_{f,\textrm{exc}}|)$. The following theorem derives a lower
bound on the ccdf of the cross-tier femtocell interference $I_{c,f}$
considering the effect of a femtocell exclusion region.

\begin{figure} [!h]
\begin{center}
   \includegraphics[width=3.5in]{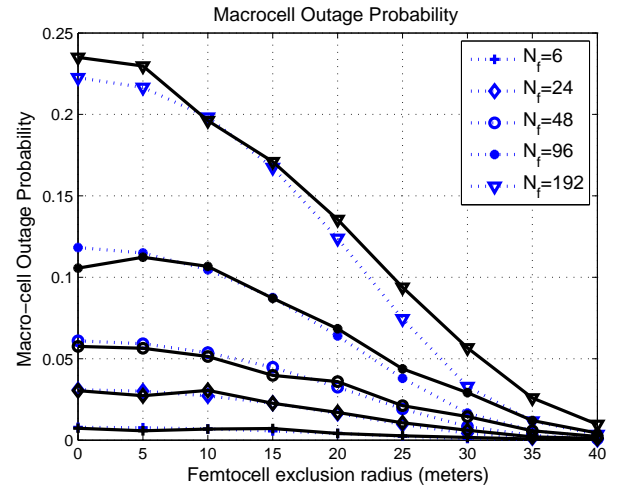}
   \caption{Macrocell outage probability for different femtocell densities with a femtocell exclusion region ($N_c=24, P_r^{f}=P_r^{c}$). Dotted lines represent the theoretical lower bounds on outage probability and solid lines represent the empirically estimated probabilities.}
   \label{fig:Pout_Mac_Femexcl}
   \end{center}
\end{figure}

\begin{lemma}[Femtocell exclusion region]
\label{Le:Mac_Femtoexcl} \emph{With a femtocell exclusion region of
radius $R_{f,\textrm{exc}}$ around the reference macrocell BS, the ccdf of
cross-tier femtocell interference $I_{c,f}$ is lower bounded as:
\begin{align}
\label{eq:Mac_Femtoexcl}
 \bar{F}_{I_{c,f}}(y) \geq 1-e^{-\pi \eta_f H(y)}
\end{align}
where $H(y)$ is defined as,
\begin{align}
H(y) &\triangleq (\frac{Q_f}{y})^{\delta}(\mathbb{E}[\Psi^{\delta}]-F_{\Psi}(u)\mathbb{E}[\Psi^{\delta}|
\Psi \leq u])-\bar{F}_{\Psi}(u)(R_{f,\textrm{exc}})^2 \notag \\
\Psi & = \sum_{i=1}^{U} \Psi_i, 10\log_{10}\Psi_i \sim
\cN(0,2\sigma_{dB}^2) \notag \\
\delta & =2/ \alpha, u =\left(\frac{y}{Q_f}\right) \cdot (R_f^{exc})^{2/\delta} \notag \\
U & \sim X|X \geq 1, X \sim \mathrm{Poisson}(U_f)
\end{align}
}
\end{lemma}
\begin{proof}
See Appendix \ref{Proof:Mac_Femtoexcl}.
\end{proof}
Fig. \ref{fig:Pout_Mac_Femexcl} depicts the macrocell outage
performance as a function of the femtocell exclusion radius,
assuming $N_c=1, P_r^{f}/P_r^{c}=1$. Notice that even a small
exclusion radius $R_f^{exc}$ results in a significant decrease in
$\mathbb{P}_{\textrm{out}}^{c}$. The implication is that a femtocell
exclusion region can increase the number of simultaneous active
femtocell transmissions, while satisfying the macrocell outage
constraint $\mathbb{P}_{\textrm{out}}^{c} \leq \epsilon$. Once again, the
close agreement between analysis and simulation shows that only the
nearby dominant femtocell interferers influence outage events at the
macrocell BS.
\begin{corollary}
\label{Co:Mac_Femtoexcl}
 \emph{With no femtocell exclusion ($R_{f,\textrm{exc}}=0$),
the ccdf of the cross-tier femtocell interference $I_{c,f}$ at a macrocell is
lower bounded as $ \bar{F}_{I_{c,f}}(y) \geq 1-e^{-\pi \eta_f
Q_f^{\delta}\mathbf{E}\lbrack \Psi^{\delta}\rbrack y^{-\delta}}$. }
\end{corollary}
Corollary \ref{Co:Mac_Femtoexcl} is the two-tier cellular network
equivalent of Theorem 3 in Weber \emph{et al.} \cite{Weber2007}, which
derives a lower bound on the outage probability for ad hoc networks
with randomized transmission and power control. Finally, this paper
considers the influence of a femtocell tier selection based handoff
policy wherein any tier 1 cellular user within the radius $R_f$
of a femtocell BS undergoes handoff to the femtocell. In essence,
the interference caused by the nearest macrocell users is mitigated, as these
users now employ power control to the femtocell BS.
\begin{lemma}
\label{Le:Thinning_TS}
 \emph{With a tier selection policy in which any user within a
radius $R_f$ of a femtocell undergoes handoff to the femtocell BS,
the intensity of tier 1 users within $\mathcal{H}$ after handoff is given as
$\lambda_{c}^{TS}(r)=\lambda_c e^{-\lambda_f \pi {R_f}^2}$
whenever $r> R_f^{exc}$, where $R_f^{exc}$ is the femtocell
exclusion radius. }
\end{lemma}
\begin{proof}
See Appendix \ref{Proof:Thinning_TS}.
\end{proof}
\begin{remark}
For small $\lambda_f$ and $r > R_f^{exc}$, a first-order Taylor
approximation shows that $\lambda_c^{TS} \approx \lambda_c \cdot
(1-\lambda_f \pi R_f^2)$. The interpretation is that tier-selection
offers marginal benefits for small femtocell sizes ($R_f \ll R_c$).
Intuitively, a small sized femtocell does not cover ``enough space''
for significant numbers of cellular users in $\Omega_c$ to
accomplish femtocell handoff. However, Theorem \ref{Th:Crossint_fem}
shows that a small femtocell size does lead to a lower uplink outage
probability.
\end{remark}
\begin{remark} The network OCs considering the effects of a femtocell
exclusion region and tier selection can be obtained by applying
Lemmas \ref{Le:Mac_Femtoexcl} and \ref{Le:Thinning_TS} in Theorems
\ref{Th:Pout_mac} and \ref{Th:Pout_fem} respectively. In doing so,
we approximate $I_{f,f}$ as a Poisson SNP whose cdf is described by
\eqref{Th:Crossint_fem}.
\end{remark}


\section{Numerical Results}
\label{Se:Numres} System parameters are given in Table
\ref{Tbl:Sysprms}. The setup consists of the region $\mathcal{H}$
surrounded by $18$ macrocell sites to consider two rings of
interferers and $2\pi/3$ sectorized antennas at each BS. In
\eqref{eq:Mac_Femtoexcl}, the statistics of the shadowing gain
$\Psi$ were empirically estimated using the MATLAB functions
\texttt{ksdensity} and \texttt{ecdf} respectively. The OCs were
analytically obtained using Theorems
\ref{Th:Crossint_fem}-\ref{Th:Pout_fem} for an outage constraint
$\epsilon=0.1$ in \eqref{eq:Twotier_OC}. The following plots compare
the OCs for a shared spectrum network with interference avoidance
against a split spectrum network with omnidirectional femtocells.

\begin{table}[ht]
\caption{System Parameters} \label{Tbl:Sysprms} \centering
\begin{tabular}{ c| c | c}
    \hline
    \textbf{Symbol} & \textbf{Description} & \textbf{Value} \\ \hline
     $\mathcal{H}$          & Region inside reference cellsite & N/A \\
     $\Omega_c,\Omega_f$ & SPPPs defining Tier 1, Tier 2 users & N/A \\
     $R_c,R_f$  & Macrocell, Femtocell Radius     & $500,20$ meters \\
     $U_f$      & Poisson mean users per femtocell & 5 \\
     $N_{\textrm{sec}}$  & Antenna sectors  & 3\\
     $N_{\textrm{hop}}$  & CDMA Hopping slots& $1,2,4$ \\
     $G$        & Processing Gain &      $128$ \\
     $\gamma$   & Target SIR per tier &      $2$ [C/I=$3$ dB] \\
     $\epsilon$ & Target Outage Probability  & 0.1 \\
    $P_r^{c}$ & Macrocell receive power & 1 \\
    $P_r^{f}$ & Femtocell receive power & 1,10,100 \\
    $\sigma_{dB}$ & Lognormal shadowing parameter & $4$ dB \\
    $\alpha,\beta$ & Path loss exponents &  $4,2$ \\
    $d_{0c},d_{0f}$ & Reference distances & $100, 5$ meters \\
    $f_c$ & Carrier Frequency & $2$ GHz \\
    \hline
\end{tabular}
\end{table}

\begin{figure} [!h]
\begin{center}
      \includegraphics[width=3.5in]{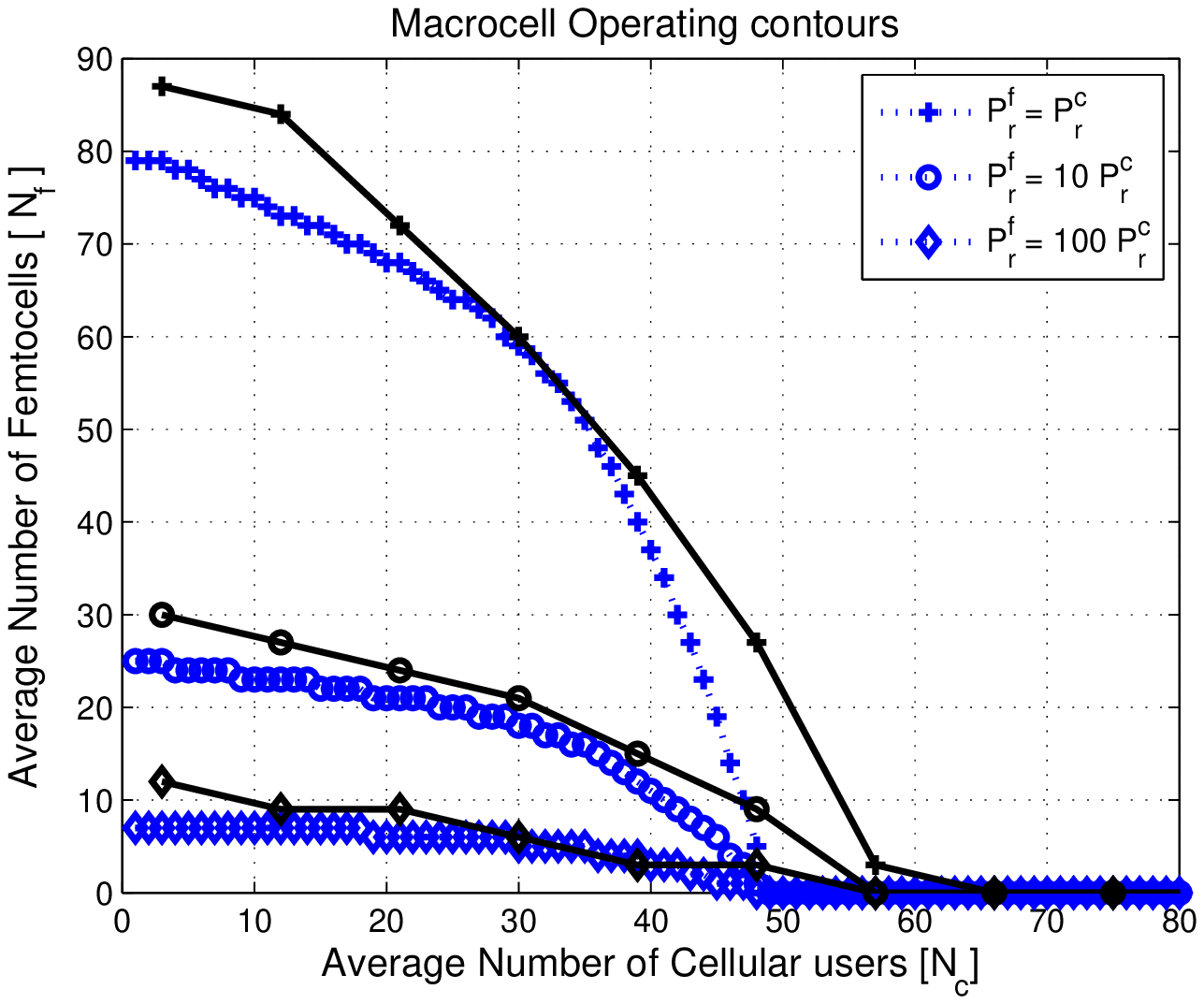}
        \caption{Operating contours for a macrocell with $N_{\textrm{hop}}=1, N_{\textrm{sec}}=3$. Solid lines show the empirical OCs, while dotted lines show the theoretically obtained OCs.}
        \label{fig:Res1}
   \end{center}
\end{figure}

\begin{figure} [!h]
\begin{center}
      \includegraphics[width=3.5in]{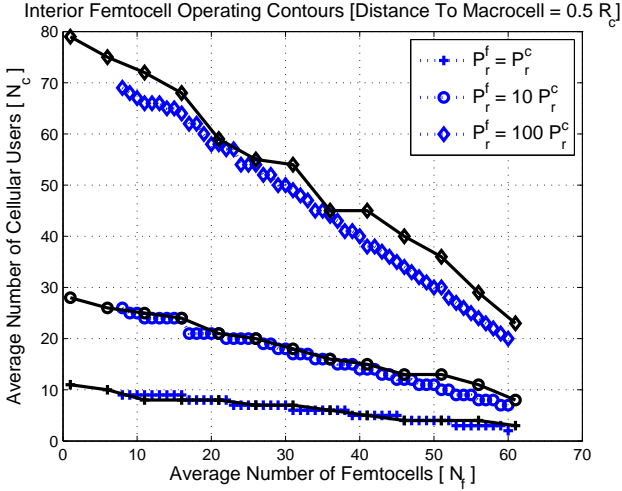}
        \caption{Operating contours for a cell interior femtocell (distance to macrocell=$0.5R_c$) with $N_{\textrm{hop}}=1, N_{\textrm{sec}}=3$. Solid lines show the empirical OCs, while dotted lines show the theoretically obtained OCs.}
         \label{fig:Res2}
   \end{center}
\end{figure}

Figs. \ref{fig:Res1} and \ref{fig:Res2} plot OCs for a macrocell and
interior femtocell with $P_r^{f}/P_r^{c}=1,10,100$ and $N_{\textrm{hop}}=1$.
The femtocell uses a sectorized receive antenna with $N_{\textrm{sec}}=3,
\theta=2 \pi/3$. The close agreement between the theoretical and
empirical OC curves indicates the accuracy of the analysis. Observe
that the outage constraints oppose one another: Relative to the macrocell, increasing
$P_r^{f}/P_r^{c}$ decreases the largest sustainable $N_f$ for
a given $N_c$. From the femtocell
standpoint, increasing $P_r^{f}/P_r^{c}$ increases the
largest sustainable $N_c$ for a given $N_f$.
\begin{figure} [!h]
\begin{center}
       \includegraphics[width=3.5in]{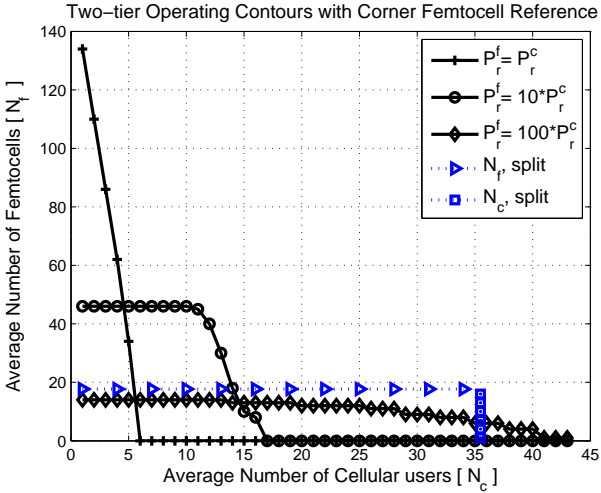}
\caption{Network operating contours for different macrocell-femtocell received power ratios and fixed hopping slots ($N_{\textrm{hop}}=4, N_{\textrm{sec}}=3$) for a cell edge femtocell (distance to macrocell $=R_c$)}
        \label{fig:Res3}
   \end{center}
\end{figure}

\begin{figure} [!h]
\begin{center}
        \includegraphics[width=3.5in]{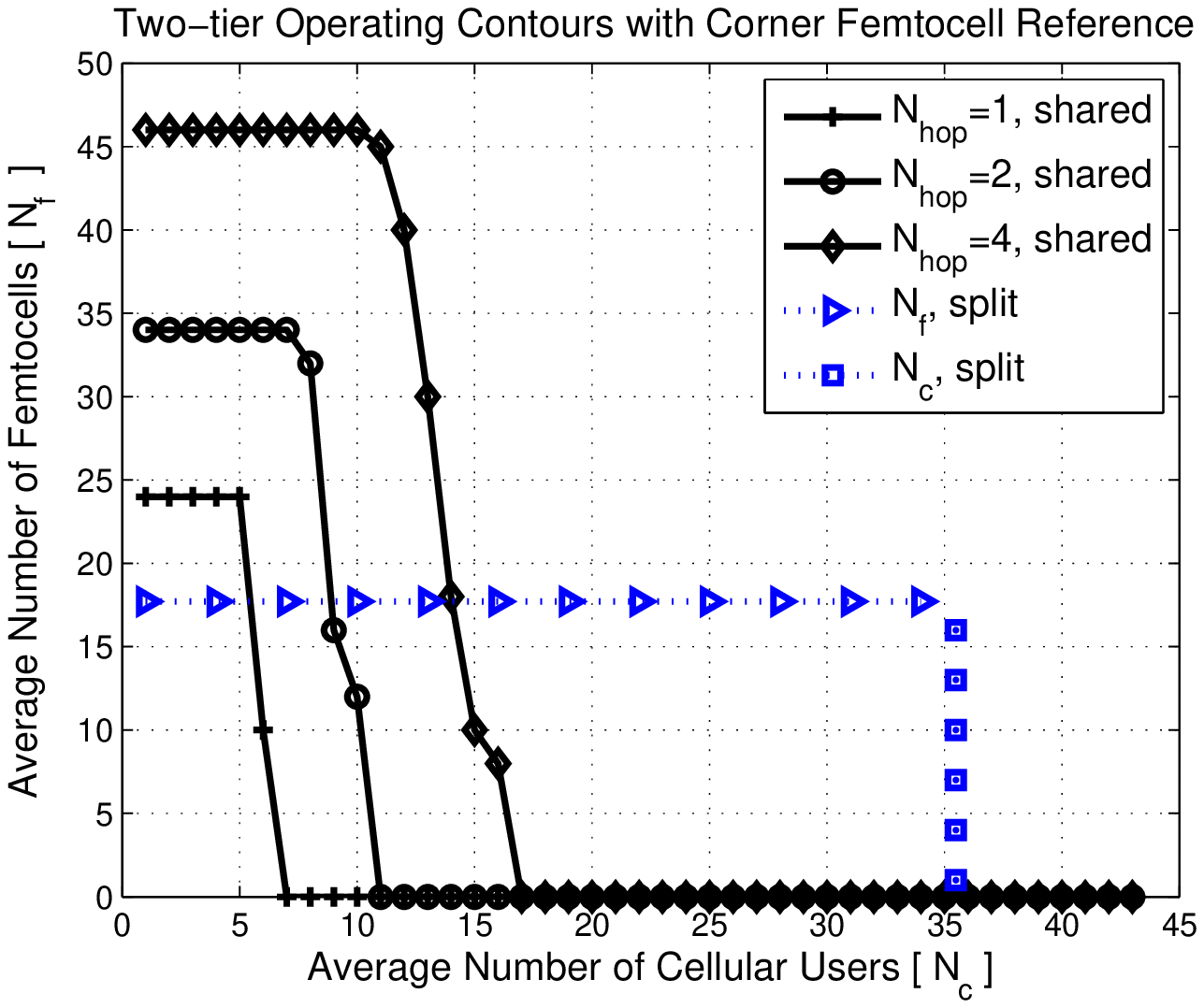}
        \caption{Network operating contours with different hopping slots ($\frac{P_r^{f}}{P_r^{c}}=10, N_{\textrm{sec}}=3$) for a cell edge femtocell (distance to macrocell $=R_c$)}
        \label{fig:Res4}
   \end{center}
\end{figure}

\begin{figure} [!h]
\begin{center}
   \includegraphics[width=3.5in]{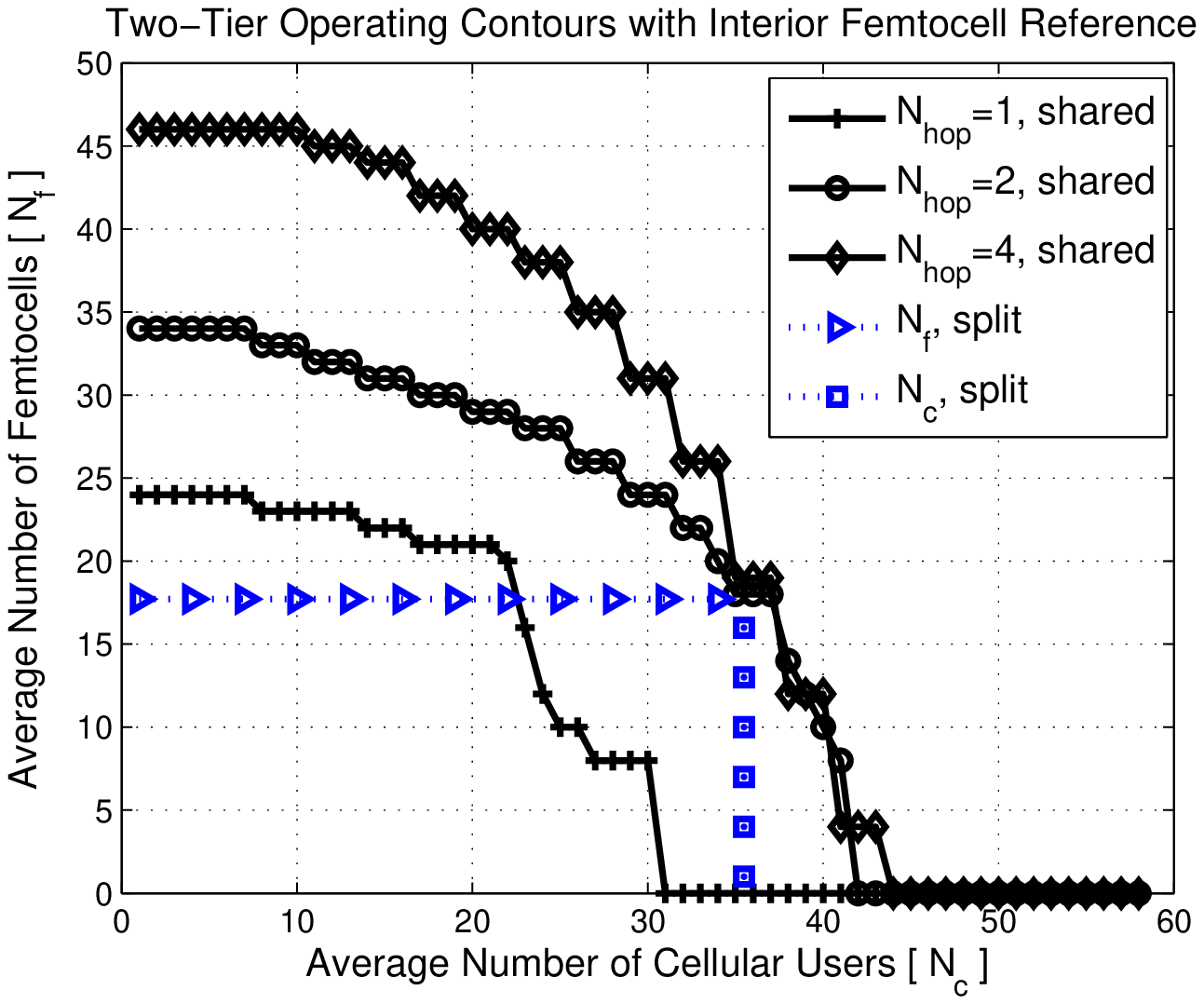}
   \caption{Network operating contours with different hopping slots ($\frac{P_r^{f}}{P_r^{c}}=10, N_{\textrm{sec}}=3$) for a cell interior femtocell (distance to macrocell $=0.5R_c$)}
   \label{fig:Res5}
   \end{center}
\end{figure}

\begin{figure} [!h]
\begin{center}
   \includegraphics[width=3.5in]{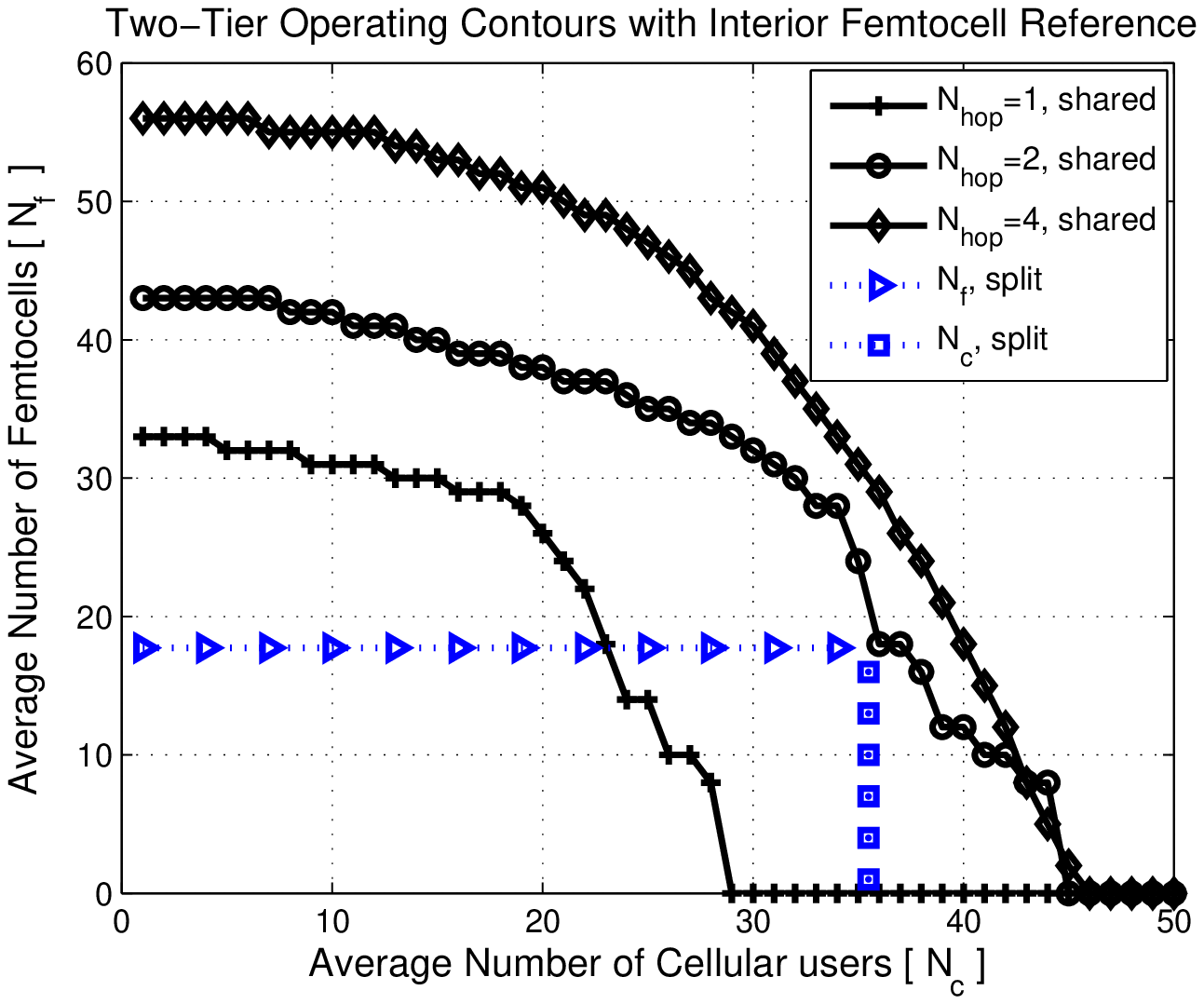}
   \caption{Network operating contours with tier selection and femtocell exclusion ($\frac{P_r^{f}}{P_r^{c}}=10, N_{\textrm{sec}}=3, R_f^{exc}=20$) for a cell interior femtocell (distance to macrocell $=0.5R_c$)}
   \label{fig:Res6}
   \end{center}
\end{figure}
Figs. \ref{fig:Res3} through \ref{fig:Res5} plot the performance of
the shared spectrum network employing interference avoidance for a
corner and an interior femtocell, as a function of $N_{\textrm{hop}}$ and
$P_r^{f}/P_r^{c}$. Fig. \ref{fig:Res3} shows that with
$P_r^{f}/P_r^{c}=1$ and a lightly loaded tier 1 network, the corner
femtocell can achieve a nearly $7$x improvement in $N_f$
relative to the split spectrum network. Intuitively, with
$P_r^{f}/P_r^{c}=1$, a macrocell BS tolerates a large cross-tier
interference; the downside being that the femtocell BS experiences higher
cellular interference arising from tier 1 users transmitting at maximum
power near the cell edge. This explains why $N_f$ decreases rapidly
with increasing $N_c$ in the OC curves for a corner femtocell. This also suggests that
achieving load balancing by increasing $N_c$ at the expense
of $N_f$ requires an order wise difference in the femtocell-macrocell receive power ratio.
We suggest that a practical wireless system use a larger femtocell-macrocell receive
power ratio ($P_r^{f}/P_r^{c}$) at the corner femtocell, relative to an
interior femtocell. Such a position based variable power ratio would ensure
that both the interior and corner femtocells can tolerate interference from
an almost identical number of tier 1 users.

With $P_r^{f}/P_r^{c}=10$ and $N_{\textrm{hop}} = 4$ slots, the OCs for corner and interior femtocells in
Figs. \ref{fig:Res4} and \ref{fig:Res5} offer greater than
$2.5$x improvement in $N_f$ relative to the split spectrum network.
Additionally, a greater degree of load balancing is achieved: with
an interior femtocell location, a maximum of $N_c=45$ tier 1 users
can be accommodated.

Fig. \ref{fig:Res6} shows the two-tier OCs when users in each tier
employ a femtocell exclusion region and a tier selection policy for
femtocell handoff. We observe an increase in $N_f$ by up to $10$
additional femtocells (or $10*U_f=50$ users) for $N_c < 30$ users.
Both femtocell exclusion and tier selection do not result in a
higher $N_c$. The reason is that a femtocell exclusion region does
not alleviate tier 1 interference at a femtocell. Furthermore, an explanation for the conservative
gains in $N_f$ is that there is a maximum tolerable interference to sustain the outage
requirements at a given femtocell, that prevents a substantial increase in the number of actively transmitting femtocells. Next, owing to small femtocell sizes, a tier selection policy succeeds in curbing tier 1
interference mainly for a large $N_f$, which is sustainable when $N_c$ is
small (to satisfy $\mathbb{P}_{out}^c \leq \epsilon$). This explains
the dominant gains in $N_f$ at a low-to-moderate $N_c$.

A relevant question is to ask: \emph{``How does the system capacity with randomly placed users and hotspots compare against a two-tier network with a given configuration?"} Due to space limitations, our paper does not address this question directly. We refer the reader to Kishore \emph{et al.} \cite[Page 1339]{Kishore2006a}. Their results show diminishing gains in the system capacity in absence of interference avoidance, because the configuration contains high levels of cross-tier interference.

Kishore proposes to alleviate cross-tier interference by varying the macrocell coverage region, through exchanging the pilot channel strength with the microcell. Our model assumes that femtocells (placed by end consumer) operate with \emph{minimal information exchange} with the macrocell BS. Due to reasons of security and scalability--there may be hundreds of embedded femtocells within a densely populated macrocell-- handing off unsubscribed users from macrocell to a femtocell hotspot may not be practical. Moreover, femtocell hotspots have a small radio range ($< 50$ meters). This necessitates an interference avoidance strategy.

\section{Conclusion}
\label{Se:Conc} This paper has presented an uplink capacity analysis
and interference avoidance strategy for a shared spectrum two-tier
DS-CDMA network. We derive exact outage probability at a macrocell
and tight lower bounds on the femtocell outage probability.
Interference avoidance through a TH-CDMA physical layer coupled with
sectorized receive antennas is shown to consistently outperform a
split spectrum two-tier network with omnidirectional femtocell
antennas. Considering the worst-case interference at a corner
femtocell, the network OCs show a 7x improvement in femtocell
density. Load balancing users in each tier is achievable through a
orderwise difference in receive powers in each tier. Additional
interference avoidance using a femtocell exclusion region and a tier
selection based femtocell handoff offers conservative improvements
in the OCs. The message is clear: Interference avoidance strategies
can make shared spectrum two-tier networks a viable proposition in
practical wireless systems.

\section*{Acknowledgement}
We acknowledge Dr. Alan Gatherer and Dr. Zukang Shen of Texas
Instruments for their valuable input and sponsorship of this
research. The contributions of Andrew Hunter are also gratefully
acknowledged.

\appendices
\section{}
\label{Proof:Thinning} Consider the Poisson field of interferers as
seen at any antenna sector (either macrocell or femtocell BS) with
antenna alignment angle $\theta$. Assuming a perfect antenna radiation pattern,
the interfering Poisson field forms heterogeneous SPPPs $\hat{\Omega}_c$ and
$\hat{\Omega}_f$ with intensities given by,
\begin{align}
\label{eq:sctrSPPP}
\hat{\lambda}_c(r,\phi)&=\frac{\lambda_c}{N_{\textrm{hop}}}\cdot\mathbf{1}(\phi \in
\lbrack \theta,\theta+\frac{2\pi}{N_{\textrm{sec}}} \rbrack) \notag \\
\hat{\lambda}_f(r,\phi)&=\frac{\lambda_f}{N_{\textrm{hop}}}(1-e^{-U_f})\cdot \mathbf{1}(\phi
\in \lbrack \theta,\theta+\frac{2\pi}{N_{\textrm{sec}}} \rbrack)
\end{align}
where $\mathbf{1}(\cdot)$ represents the indicator function. The
following observations rigorously explain \eqref{eq:sctrSPPP}.
\begin{asparadesc}
\item[Hopping slot selection:] The set of macrocell users and femtocell BSs transmitting over any
hopping slot is obtained by independent Bernoulli thinning of the
SPPPs $(\Omega_c,\Omega_f)$ by the probability of choosing that
hopping slot namely $1/N_{\textrm{hop}}$.
\item[Active femtocell selection:] The factor $(1-e^{-U_f})$ arises
because the set of femtocells with \emph{at least one actively
transmitting user} is obtained using independent Bernoulli thinning
of $\Omega_{f}$ \cite{Kingman}. Observe that a femtocell with $U
\geq 1$ actively transmitting users satisfies
$\mathbb{E}[U]=\frac{U_f}{1-e^{-U_f}}$.
\end{asparadesc}
The event consisting of marking femtocells by the probability that
they contain at least one actively transmitting user and the event
of marking femtocells by the probability of choosing a common
hopping slot are independent; this implies that the resulting SPPP
$\hat{\Omega}_f$ has intensity $\frac{\lambda_f}{N_{\textrm{hop}}}
(1-e^{-U_f})$. Finally, using the Mapping theorem \cite[Section 2.3]
{Kingman} for Poisson processes, one can map the
heterogeneous SPPPs $\hat{\Omega}_c$ and $\hat{\Omega}_f$ over one
antenna sector to homogeneous SPPPs $\Phi_c$ and $\Phi_f$ over
$\mathbb{R}^2$ with intensities
$\eta_c=\frac{\lambda_c}{N_{\textrm{hop}}N_{\textrm{sec}}}$ and
$\eta_f=\frac{\lambda_f}{N_{\textrm{hop}}N_{\textrm{sec}}}(1-e^{-U_f})$
respectively.

\section{}
\label{Proof:Crossint_fem} From \eqref{eq:PoisSNP}, $I_{c,f}$ (and
$I_{f,f}$) are distributed as a Poisson SNP $\hat{Y}=\sum_{i \in
\hat{\Omega}_{f}}Q_f \Psi_i |X_i|^{-\alpha}$ over an antenna sector
of width $2 \pi/N_{\textrm{sec}}$. Next, the Mapping theorem \cite{Kingman}
is used to prove \eqref{eq:DF1DSPPP}.
\begin{asparaenum}
\item Invoke Lemma \ref{Le:Thinning} for mapping
$\hat{\Omega}_f$ to a homogeneous SPPP $\Phi_f$ on $\mathbb{R}^2$.
This implies that $\hat{Y}$ is distributed identically as $Y=
\sum_{i \in \Phi_{f}}Q_f \Psi_i |X_i|^{-\alpha}$.
\item Map the planar SPPP defining $\Phi_{f}$ with intensity $\eta_f$
to a 1D SPPP with intensity $\pi \eta_f$ using Proposition 1,
Theorem 2 in \cite{Ilow1998}. For doing so, rewrite $Y$ as,
$Y=\sum_{i \in \Phi_{f}}Q_f \Psi_i ({|X_i|^2})^{-\alpha/2}$ which
represents a SPPP on the line with Poisson arrival times $|X_i|^2$
and intensity $\pi \eta_f= \frac{\pi \lambda_f}{N_{\textrm{hop}}\cdot
N_{\textrm{sec}}}(1-e^{-U_f})$.
\end{asparaenum}
Consequently, $Y$ is identically distributed as a 1D SPPP with
intensity $\pi \eta_f$, which represents a L\'{e}vy-stable
distribution with stability exponent $\delta=2/\alpha$ \cite{Taqqu},
and a characteristic function given by $Q_{Y}(s)=\exp{\lbrack -\pi
\eta_f \Gamma(1-\delta) \mathbf{E}\lbrack \Psi^{\delta} \rbrack (Q_f
s)^{\delta}\rbrack}$, where $\Gamma(z) \triangleq
\int_{0}^{\infty}t^{z-1}e^{-t}dt$ is the gamma function. In
particular, when $\alpha=4$, $Y$ follows a L\'{e}vy-stable
distribution with stability exponent $\delta=0.5$, with statistics
\eqref{eq:DF1DSPPP} obtained from Equation (30) in \cite{Lowen1990}.

\section{}
\label{Proof:Pout_mac} At the macrocell BS, the interference denoted
by $I_{c,\textrm{in}},I_{c,\textrm{out}}$ and $I_{c,f}$ are mutually independent
random variables. The macrocell outage probability
$\mathbb{P}_{\textrm{out}}^{c}$ defined in \eqref{eq:Outg_const} can be
computed by the probability of the complementary event,
corresponding to the probability that the cumulative interference
does not exceed the SIR threshold $\rho_c=G P_r^{c}/(\gamma N_{\textrm{hop}})$.
The cdf of $(I_{c,\textrm{in}}+I_{c,\textrm{out}}+I_{c,f})$ can be computed
using a three-fold convolution. Observe that the event that the
intra-tier macrocell interference from $(k-1)$ in-cell tier 1 interferers
$I_{c,\textrm{in}}$ equals $(k-1)P_{r}^{c}$, given at least one active
tier 1 user (user of interest), is equivalent to the event that
$\Phi_{c}$ has exactly $k$ elements within $\mathcal{H}$.
The probability of this event is given as,
\begin{align}
\mathbb{P} \lbrack I_{c,\textrm{in}}&=(k-1) \cdot P_r^{c} \mid k \geq 1 \rbrack \notag \\
&=\mathbb{P} \lbrack \lvert \Phi_{c} \rvert=k \mid \lvert \Phi_{c}\rvert
\geq 1 \rbrack \notag \\
&=\frac{1}{1-e^{-\eta_c |\mathcal{H}|}}\frac{e^{-\eta_c |\mathcal{H}|}{(\eta_c
|\mathcal{H}|)^k}}{k!}
\end{align}
The total interference caused by the $(k-1)$ interfering macrocell
users equals $(k-1)\cdot P_r^{c}$. Outage does not occur if the
residual interference $I_{c,\textrm{out}}+I_{c,f}$ is less than $\rho_c-(k-1)P_r^{c}$.
Using Theorem \ref{Th:Crossint_fem} and independence of the
Gaussian distributed $I_{c,\textrm{out}}$ and $I_{c,f}$, the result follows.

\section{}
\label{Proof:Crossint_mac} The interference experienced at a
femtocell antenna sector $\theta \leq \phi \leq
\theta+2\pi/N_{\textrm{sec}}$ is lower bounded by the cellular interference
arising within $\mathcal{H}_{sec}$. If the femtocell BS is located at
distance $R_0$ from the reference macrocell, then any macrocell user
located at polar coordinates $(r_i,\phi_i)$ w.r.t the femtocell BS
causes an interference equaling $P_r^{c}\Psi_i(|R_0+re^{\mathbf{i} \phi}|/r)^\alpha$
at the femtocell BS. Corresponding to the heterogeneous SPPP $\Pi_c$
(see Def. \ref{De:def3}), outage events at the femtocell BS
arising from cellular interference $I_{f,c}$ can be categorized into two
types: In the first type, outage events arise due to interference caused by a
single user in $\Pi_{c}$. The second class of outage events occur
due to the macrocell interferers whose \emph{cumulative} interference causes
outage \cite{Weber2005}. This class precludes all interferers
falling in the first category. Mathematically, for an outage
threshold $y$ at the femtocell BS, split $\Pi_{c}$ into two disjoint
heterogeneous Poisson SPPPs $\Pi_{c}=\Pi_{c,y} \cup \Pi_{c,y}^{C}$
corresponding to the set of \emph{dominant} and \emph{non-dominant}
cellular interferers:
\begin{align}
\label{eq:SPPPartng}
\Pi_{c,y} &\triangleq \{(r_i,\phi_i) \in
{\Pi}_{c} : P_r^{c}\Psi_i (|r_{i}e^{\mathbf{i} \phi_i}+R_0|/r_i)^{\alpha} \geq
y \} \notag \\
\Pi_{c,y}^{C}&=\Pi_c \setminus \Pi_{c,y}
\end{align}
At any point $(r,\phi) \in \mathcal{H}$, the intensity of $\Pi_{c,y}$
denoted by $\lambda_{c,y}(r,\phi)$ is given as,
\begin{align}
\label{eq:InhomoSPPP} \lambda_{c,y}(r,\phi)=
\frac{\lambda_c}{N_{\textrm{hop}}}\bar{F}_{\Psi}\Biggl \lbrack
\frac{yr^{\alpha}}{P_r^{c}|r e^{\mathbf{i} \phi}+R_0|^{\alpha}} \Biggr \rbrack
\cdot \mathbf{1}(\phi \in [\theta,\theta+\frac{2\pi}{N_{\textrm{sec}}}])
\end{align}
In the event of $\Pi_{c,y}$  being non empty, the femtocell BS
experiences outage, arising from the interference caused by a user in
$\Pi_{c,y}$. Therefore, $\mathbb{P}_{\textrm{out}}^{f}$ is lower bounded by
the probability that $\Pi_{c,y}$ has at least one element. Equation
\eqref{eq:Pout_sctrdintfemlb} results from the Poisson void
probability of the complementary event $\mathbb{P}{[|\Pi_{c,y}|=0]}$
\cite{Kingman}. This completes the proof for the first assertion.

To prove \eqref{eq:Pout_omnicorfemlb}, recognize that a corner
femtocell with an omnidirectional BS antenna encounters
cellular interference from the three surrounding cellsites. The dominant
macrocell interferer set $\Pi_{c,y}$ can be expressed as
$\Pi_{c,y}=\bigcup_{i=1}^{3}\Pi_{c,y}^{i}$, where $\Pi_{c,y}^{i}$
denotes the dominant macrocell interferer set in neighboring
cellsite $i$. The heterogeneous SPPPs $\Pi_{c,y}^{i}$ are
non-intersecting with an intensity expressed by
\eqref{eq:InhomoSPPP}. The ccdf of $I_{f,c}$ is then lower bounded
by the probability of $\Pi_{c,y}$ being non empty, which can be
deduced from the event that $\Pi_{c,y}^{i},i \in \lbrace 1,2,3 \rbrace$ are empty.
\begin{align}
\label{eq:femtointerferencelb}
F^{lb}_{I_{f,c}}(y)&=\prod_{i=1}^{3} \mathbb{P}{(|\Pi_{c,y}^{i}|=0)} \notag \\
                  &=\exp \left\{-3\frac{\lambda_c}{N_{\textrm{hop}}} \iint\limits_{\mathcal{H}}S(r,\phi;y) r
\ \textrm{d}r \textrm{d}\phi \right\}
\end{align}
To complete the proof, use pairwise independence of the events that
$\Pi_{c,y}^{i}$ and $\Pi_{c,y}^{j}$ are empty and $S(r,\phi;y)$ in
\eqref{eq:Pout_sctrdintfemlb} to show that
$\bar{F}_{I_{f,c}}(\cdot)$ is lower bounded as $\bar{F}_{I_{f,c}}(y)
\geq 1-{F}^{lb}_{I_{f,c}}(y)$ in \eqref{eq:femtointerferencelb}.

\section{}
\label{Proof:Pout_fem} The number of femtocell users within a
femtocell antenna sector is Poisson distributed with mean
$U_f/N_{\textrm{sec}}$. The overall interference is composed of three terms namely
$I_{f,\textrm{in}},I_{f,f}$ and $I_{f,c}$ which are mutually independent.
Given $m$ actively transmitting femtocell users including the user
of interest, the interference from users within the femtocell equals $I_{f,\textrm{in}}=(m-1)P_r^{f}$.
The threshold for $I_{f,f}+I_{f,c}$ to cause outage therefore equals
$\tilde{\rho}_f=\rho_{f}-(m-1)P_r^{f}, \rho_f \triangleq G
P_r^{f}/(\gamma N_{\textrm{hop}})$ using \eqref{eq:Outg_const}. A lower
bound on $\mathbb{P}_{\textrm{out}}^{f}$ is obtained as,
\begin{align}
1&-\mathbb{P}_{out}^{f,lb} \notag \\
           \label{eq:Pout_fem_step1}
           &=\frac{e^{-U_{f,\textrm{sec}}}}{1-e^{-U_{f,\textrm{sec}}}}\sum_{m=1}^{\lfloor
           \rho_f/P_r^{f} \rfloor}\frac{U_{f,\textrm{sec}}^m}{m!} \cdot F_{I_{f,c}^{lb}+I_{f,f}} (\tilde{\rho}_{f}) \\
           &\overset{(a)}=\frac{e^{-U_{f,\textrm{sec}}}}{1-e^{-U_{f,\textrm{sec}}}}\sum_{m=1}^{\lfloor
           \rho_f/P_r^{f} \rfloor}\frac{U_{f,\textrm{sec}}^m}{m!}\cdot [F_{I_{f,c}}^{lb}\ast f_{I_{f,f}}](\tilde{\rho}_{f}) \notag \\
           &\overset{(b)}\approx \frac{e^{-U_{f,\textrm{sec}}}}{1-e^{-U_{f,\textrm{sec}}}}\sum_{m=1}^{\lfloor
           \rho_f/P_r^{f}\rfloor}\frac{U_{f,\textrm{sec}}^m}{m!} \cdot
           [(1+\ln(F_{I_{f,c}}^{lb}))\ast f_{I_{f,f}}](\tilde{\rho}_f) \notag
\end{align}
Equation \eqref{eq:Pout_fem_step1} uses the lower bound on macrocell
interference $I_{f,c}^{lb}$ arising from the set of dominant macrocell
interferers \eqref{eq:SPPPartng}. Step (a) uses
pairwise independence of $I_{f,f}$ and $I_{f,c}$ for performing a
convolution of the respective probabilities. Finally,
Step (b) follows from a first-order Taylor series
approximation of $F_{I_{f,c}}^{lb}$ in \eqref{eq:Pout_sctrdintfemlb}
using $e^{x} \approx (1+x)$ for small $\lambda_c$ in the low outage
regime.

\section{}
\label{Proof:Mac_Femtoexcl} Outside the femtocell exclusion
region ${\cR}_{f}^{exc} \subset \mathcal{H}$, corresponding to an outage threshold $y$, the SPPP
$\Phi_f$ (see Def.\ref{De:def2}) of intensity $\eta_f=\frac{\lambda_f}{N_{\textrm{hop}}N_{\textrm{sec}}}(1-e^{-U_f})$
can be split into the dominant and non-dominant interfering femtocells denoted by
$(\Phi_{f,y},\Phi_{f,y}^{C})$ respectively. The
heterogeneous SPPP $\Phi_{f,y}=\{(r_i,\phi_i) \in
\Phi_{f} : Q_f\Psi_{i} r_{i}^{-\alpha} \geq y \}$ consists
of actively transmitting femtocells over $\mathbb{R}^2$ which are capable of
individually causing outage at a macrocell BS. At any $(r,\phi)$
w.r.t macrocell BS, the intensity of $\phi_{f,y}$ equals
$\eta_f \cdot \bar{F}_{\Psi}(y r^{\alpha}/{Q_f})$. The ccdf of the femtocell interference
$I_{f,c}$ is lower bounded by the probability that
$\Phi_{f,y}$ is non-empty. For if $\Phi_{f,y}$
contains at least one element, then the macrocell antenna sector
is in outage (by construction of $\Phi_{f,y}$). Using the
void probability of $\Phi_{f,y}$, the lower bound $\bar{F}^{lb}_{I_{c,f}}(y)$
is given as,
\begin{align}
1&-\bar{F}^{lb}_{I_{c,f}}(y) \notag \\
\label{eq:Mac_Femtoexcl_step1}
&= \exp
\left\{-2\pi\eta_f
\int \limits_{R_{f,\textrm{exc}}}^{\infty}\bar{F}_{\Psi}\left(\frac{y r^{\alpha}}{Q_f} \right)  r \textrm{d}r \right\} \\
&\overset{(a)}=\exp \left \{-\pi \eta_f
Q_f^{\delta}y^{-\delta} \int \limits_{u}^{\infty}
\bar{F}_{\Psi}(t) d(t^{\delta})\right \} \notag \\
&\overset{(b)}=\exp \left \{-\pi \eta_f
Q_f^{\delta}y^{-\delta} \Biggl \lbrack \int \limits_{u}^{\infty} t^{\delta}f_{\Psi}(t) dt-
\bar{F}_{\Psi}(u)(R_{f,\textrm{exc}})^2 \Biggr \rbrack \right \} \notag
\end{align}
Step (a) in \eqref{eq:Mac_Femtoexcl_step1} follows by substituting $t=y
r^{\alpha}/Q_f$ and $\delta=2/ \alpha$ in \eqref{eq:Mac_Femtoexcl_step1}, while
Step (b) is obtained using integration by
parts and setting $u=(\frac{y}{Q_f})(R_{f,\textrm{exc}})^{2/ \delta}$.
Using $\int_{u}^{\infty}t^{\delta}f_{\Psi}(t)dt=
\mathbb{E}[\Psi^{\delta}]-F_{\Psi}(u)\mathbb{E}[\Psi^{\delta}|
\Psi \leq u]$ in Step (b) completes the proof.

\section{}
\label{Proof:Thinning_TS} In the region $0 \leq r \leq R_{f,\textrm{exc}}$
around the reference macrocell, actively transmitting femtocells are
absent, so that there are no femtocells for handoff to occur for any
user in $\Omega_c$. Consequently, the intensity of the tier 1
cellular users in $0 < r < R_{f,\textrm{exc}}$ equals $\lambda_c$. For
$r > R_{f,\textrm{exc}}$, the intensity of the macrocell users is found by
computing the probability that any point in $\Omega_c$ (prior tier
selection) does not fall within $R_f$ meters of a femtocell BS. This
is equivalent to computing the void probability of $\Omega_f$ within
a circle of radius $R_f$ of every point in $\Omega_c$, which equals
$e^{-\lambda_f \pi R_f^2}$.

This paper assumes an independent Bernoulli thinning of each
point in $\Omega_c$ by the probability that a tier 1 user falls with
$R_f$ of a femtocell. Strictly speaking, this statement is not
correct: Given two closely spaced tier 1 users in $\Omega_c$, the
event that the first user undergoes femtocell handoff is correlated
with a nearby user in $\Omega_c$ undergoing handoff with the same
femtocell. However, we justify that this assumption is reasonable
while considering the small size of each femtocell. Then, the
intensity of tier 1 users following the femtocell handoff is
obtained by independent Bernoulli thinning of $\Omega_c$ by the void
probability $e^{-\lambda_f \pi R_f^2}$ \cite{Kingman}, which
completes the proof.

\end{document}